\newcommand{\be}{\begin{equation}}
\newcommand{\en}{\end{equation}}
\newcommand{\bea}{\begin{eqnarray}}
\newcommand{\ena}{\end{eqnarray}}
\newcommand{\nn}{\nonumber\\}
\newcommand{\imag}{\mathop{\rm Im}\nolimits}
\newcommand{\real}{\mathop{\rm Re}\nolimits}
\documentclass{webofc}
\usepackage[varg]{txfonts}   
\begin{document}
\title{\vspace{-24pt}\hbox to0pt{\vbox to24pt{\hbox to\hsize{\rm\hfill
MITP/18-087}\vfill}\hss}NLO and NNLO corrections to polarized top quark decays}
%
%

\author{\firstname{Stefan } \lastname{Groote}\inst{1}
  \fnsep\thanks{\email{groote@ut.ee}} \and
  \firstname{J\"urgen G.} \lastname{K\"orner}\inst{2}\fnsep\thanks{
    \email{jukoerne@uni-mainz.de}} 
}

\institute{F\"u\"usika Instituut, Tartu \"Ulikool,
  W.~Ostwaldi 1, EE-50411 Tartu, Estonia 
\and
     PRISMA Cluster of Excellence, Institut f\"ur Physik, Johannes-Gutenberg-Universit\"at, D-55099 Mainz, Germany     }

\abstract{%
  We present partial results on NLO and NNLO QCD, and NLO electroweak
  corrections to polarized top quark decays. In parallel we derive positivity
  bounds for the polarized structure functions in polarized top quark decays
  and check them against the
  perturbative corrections to the structure functions.
}
\maketitle
\section{Introduction}
\label{intro}
In the limited space available to us in this write-up of a talk given
at the International Workshop on QCD Theory and Experiment
(QCD@Work 2018) in Matera, Italy,
we cannot review the subject of polarized
top quark decays in any depth. Instead we take the opportunity to report on
results on radiative corrections to polarized top quark decays obtained by our 
group in the last few years. We will share our insights into the problem,
why we did the calculations and how we did them. We also take the opportunity
to specify which perturbative calculations have been done and which remain to
be done.

The motivation for studying polarized top quark decays is provided by
the huge sample of singly produced polarized top quarks at the LHC.
The dominant source of polarized top quarks is from weak $t$--channel
production with an average polarization of $\approx$ 90 $\%$ for the
produced top quarks. Up to date
the LHC detectors have seen $\sim$ $10^7$ singly produced top quarks.
The projected overall luminosity of the future high luminosity HL-LHC is
3 $ab^{-1}$ which
corresponds to $\sim\,10^9$ singly produced top quarks. Top quarks retain
their polarization at birth when they decay since the life time of the top
quark is so short. 
\section{The polarized top quark three-body decay
  $t(\uparrow) \to X_b+\ell^+ +\nu_\ell$}
\label{sec-2}
\begin{figure}[h]
\centering
\includegraphics[width=8cm,clip]{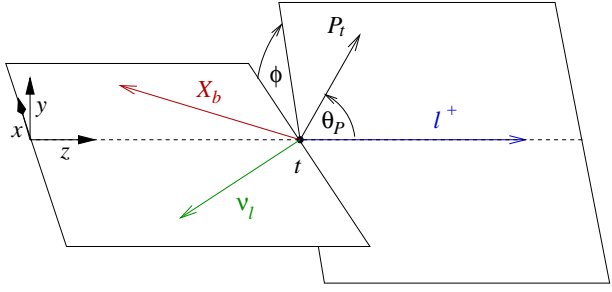}
\caption{Definition of the polar angle $\theta_P$
  and the azimuthal angle $\phi$ in the three-body decay of a polarized top quark.}
\label{fig-azi}       
\end{figure}

The full angular decay distribution for polarized top quark decay
$t(\uparrow)\to X_b+\ell^+ +\nu_\ell$ in the top quark rest frame can be written
in terms of the four structure functions $A,B,C$ and $D$. The decay
distributuion reads~\cite{Groote:2017pvc}
\bea
\label{angdis}
\frac{d\Gamma}{d\cos\theta d\phi} 
&=& A+B\,P_t\cos\theta_P+C\,P_t\sin\theta_P\cos\phi+D\,P_t\sin\theta_P\sin\phi \nn
&=&A \,\Big(1 + \frac{B}{A}\,P_t\cos\theta_P+\frac{C}{A}P_t\sin\theta_P\cos\phi
+ \frac{D}{A}P_t\sin\theta_P\sin\phi\Big)\, ,
\ena
where the angles $\theta_p$ and $\phi$ are defined in Fig.~\ref{fig-azi}. In the
classification of Ref.~\cite{Korner:1998nc} this is the helicity system Ib.

In the usual classification the structure functions $A,\,B$ and $C$ are
$T$--even structure functions and $D$ is a $T$--odd structure function as can
be seen by rewriting the angular factor multiplying $D$ in the form
\be
\sin\theta_P \sin \phi \propto \vec p_\nu\cdot(\vec p_\ell \times \vec s_t)\,.
\en
The $T$--odd structure function $D$ can be fed by final-state interactions
(also called rescattering corrections)
or by $CP$--violating interactions. We will present examples of both
contributions further on.

It is clear that the angular decay distribution must remain positive definite
over all of phase space. The positivity of the rate will be a recurring
theme in this write-up.
\subsection{The LO angular distribution}
\label{2.1}
For the leading order (LO) Born term contribution one obtains the amplitude
\bea
\label{fierz}
M &=&\bar u(b)\gamma^\mu(1-\gamma_5)u(t)\, \bar u(\ell)\gamma_\mu
(1-\gamma_5)v(\nu) \nn
 &=&2\bar u(b)(1+\gamma_5)v(\nu)\,\bar u(\ell)(1-\gamma_5) u(t) \,,
\ena
where we have used a Fierz transformation of the second kind to convert
the $(V-A)^\mu(V-A)_\mu$ form to a $(S+P)(S-P)$ form
(see e.g. \cite{Groote:2006kq}).
The polarized angular decay distribution then reads
(we set $P_t=|\vec P_t|=1$)
\bea
W^P(\cos\theta_P)&=& \sum_{spins} |M|^2\,=\,8\,tr
\{p\!\!\!/_b   p\!\!\!/_\nu\}\, tr\{(p\!\!\!/_t+m_t)
\tfrac 12(1+\gamma_5 s\!\!/_t)(1+\gamma_5)p\!\!\!/_\ell\} \nn
                &=& 16 m_t^4\,x_\ell(1-x_\ell)(1+ \cos\theta_P)\, ,
\ena
where $x_\ell=2E_\ell/m_t$.
Quite naturally, the same result is obtained more tediously if one uses the
$(V-A)^\mu(V-A)_\mu$ form of the amplitude~(\ref{fierz}).
At LO there are no azimuthal correlations, i.e. $C=D=0$! It is not
difficult to see that the abscence of LO
azimuthal correlations is in line with the postulate of positivity for
the LO rate.
\subsection{The NLO angular distribution}
\label{sec-2.2}
The NLO QCD contribution to the structure functions $A,\,B$ and $C$ have been
calculated in~\cite{Groote:2006kq}. We denote the NLO contribution by
$A^{NLO}=A^{(0)}+A^{(1)}$
etc..
Setting $\phi=0$ the angular decay distribution at NLO reads
 \be
W(\theta_P)=\,
A^{(0)} \,\Bigg(\Big(1 + \frac{A^{(1)}}{A^{(0)}}\Big)
  +\Big(1 + \frac{B^{(1)}}{A^{(0)}}\Big)\,\cos\theta_P +
    \frac{C^{(1)}}{A^{(0)}}\sin\theta_P\,\Bigg)
    \en
    where $A^{(1)}/A^{(0)}=-0.0846955$, $B^{(1)}/A^{(0)} =-0.0863048$ and
    $C^{(1)}/A^{(0)}=-0.0024$. The above values of the coefficient functions
    represent average
    values of the respective functions averaged over $x_\ell$ in the interval
    $[x,1]$ where $x=m_W/m_t$. As concerns the $\cos\theta_P$ dependence,
    the NLO rate remains barely positive for $\cos\theta_P=-1$ and $P_t=1$
    as can be seen from
    \be
    W(\theta_P)\sim 1+\frac{1+B^{(1)}/A^{(0)}}{1+A^{(1)}/A^{(0)}}
    \cos\theta_P
    =1+0.99824 \cos\theta_P \, .
    \en
    Next we analyze positivity including the $T$--even azimuthal correlation
    proportional to the strucrture function $C$. We expand
    the angular rate around the point of risk $\theta_P=\pi$.
    One has
    $\cos(\pi-\delta)=-1+\frac{1}{2}\delta$ and $\sin\delta=\delta$.
    We then obtain ($\Delta=(A^{NLO}-B^{NLO})/A^{NLO}\,=0.001758$)
    \be
    W(\theta_P)= \Delta-\frac{C^{(1)}}{A^{(0)}}
    \frac{1}{1+A^{(1)}/A^{(0)}}\,\delta + \frac{(1-\Delta)}{2}\,\delta^2 \,.
    \en
    A lower bound on the positivity of the rate is obtained when the
    discriminant of the quadratic equation vanishes. One obtains
    \be
    \label{boundC}
    \Bigg|\frac{C^{(1)}}{A^{(0)}}\Bigg|\le \underbrace{ \sqrt{2\Delta(1-\Delta)}
    (1+\frac{A^{(1)}}{A^{(0)}})}_{0.05422}\, .
    \en
    It is apparent that the NLO contributions listed above
    $(|C^{(1)}/A^{(0)}|=0.0024)$ easily satisfy the
    bound. The
    lower bound occurs at $\delta=-8.36 \times 10^{-4}\,\pi$ which shows that
    the small-$\delta$ expansion is well justified.
    This calculation provides the setting for deriving a bound
    on the size of the $T$--odd structure function $D$ to be discussed
    in the next subsection.
\subsection{NLO positivity bounds for the $T$--odd structure function $D$ }
\label{sec-2.3}
\begin{figure}[h]
    \centering
    \includegraphics[width=3cm,clip]{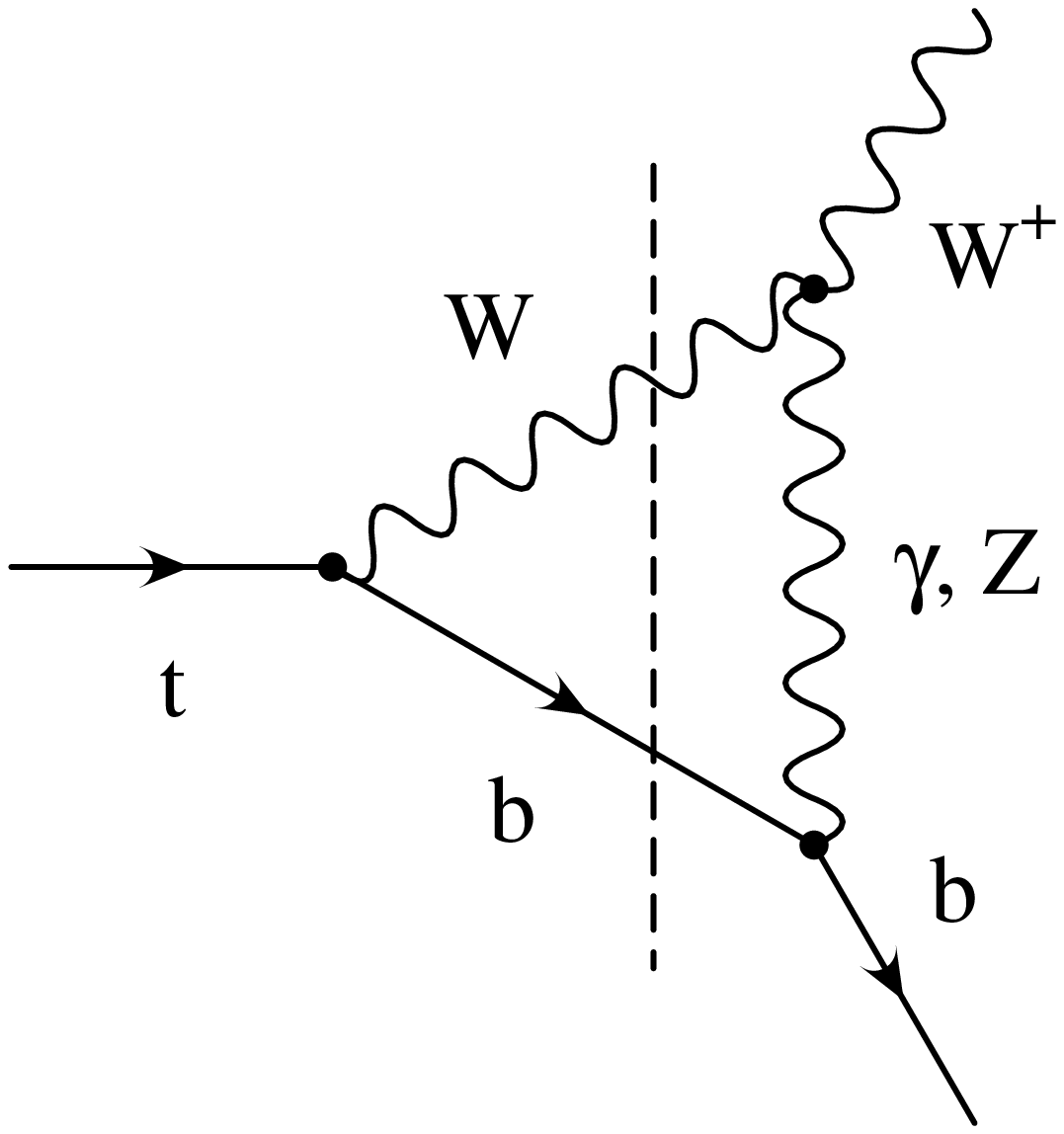} \qquad
    \includegraphics[width=3cm,clip]{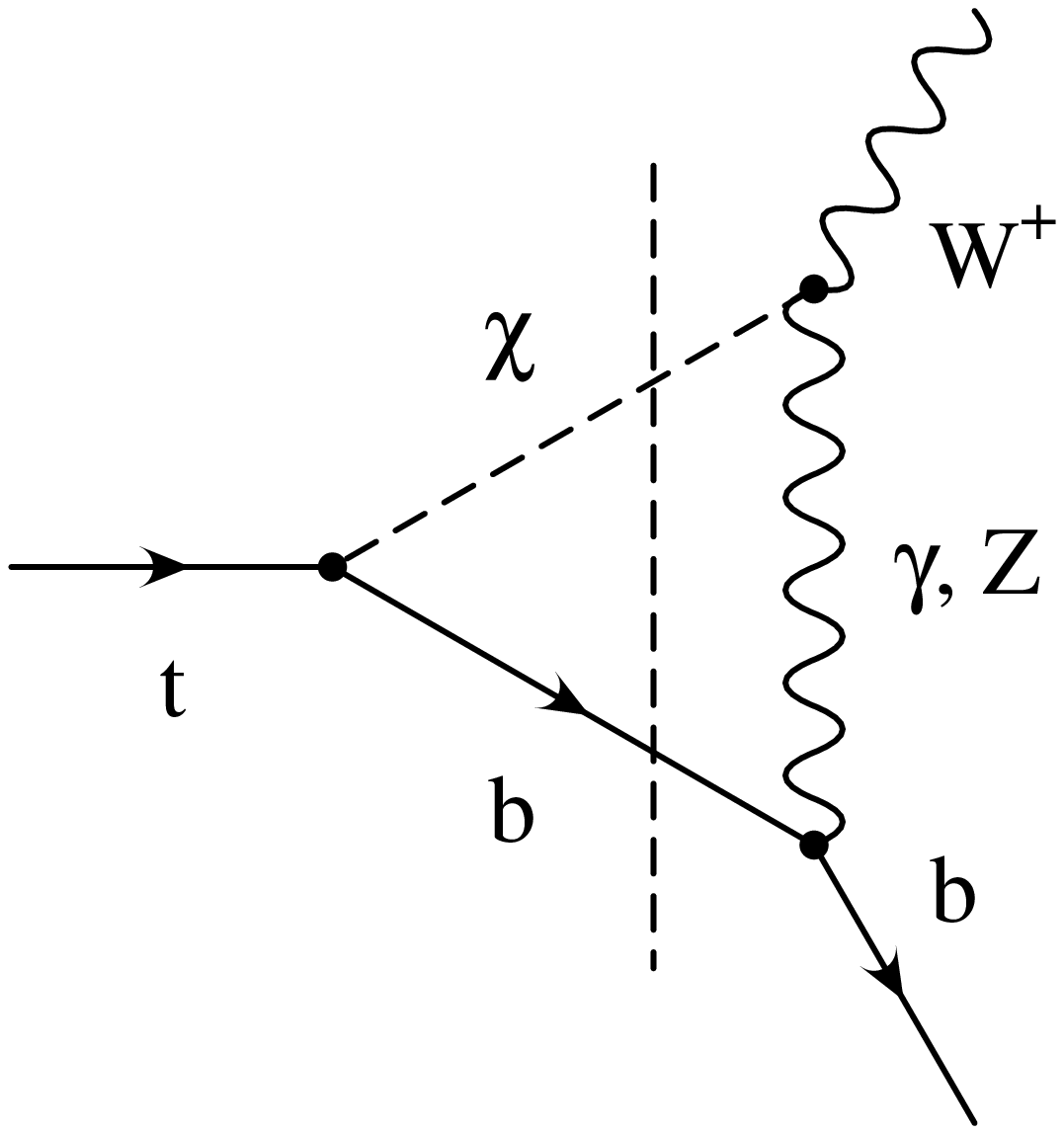}
\caption{Electroweak absorptive parts of the four Feynman diagrams that
  contribute to $T$-odd correlations in polarized top quark decays}
\label{absorp}       
  \end{figure}

There is a $CP$--conserving Standard Model contribution to the $T$--odd
structure function $D$ coming from the NLO electroweak rescattering
correction as shown as absorptive parts in Fig.~\ref{absorp}. One can check
that there are no NLO absorptive QCD contributions.

We present the result
of calculating the absorptive parts of the two diagrams in terms of the
imaginary part of the effective coupling constant $g_R$ in
the effective Lagrangian
\be
  {\cal J}^\mu_{\rm eff}=-\frac{g_w}{\sqrt{2}}\, \bar b \,
  \Big\{\gamma_\mu(V_LP_L+V_RP_R)+
  \frac{i\sigma^{\mu\nu}q_\nu}{m_W}(g_L P_L+g_R P_R)\Big\} \,t
  \en
where $P_{L,R}=(1\mp\gamma_5)/2$. The SM structure of the $tbW^+$ vertex is
obtained by dropping all terms except for the contribution proportional to
$V^\ast_{tb}\sim1$. $\imag g_R$ is contributed to by $CP$--conserving
rescattering effects and by $CP$--violating New Physics effects. 

The electroweak rescattering contribution to $\imag g_R$ can be calculated
to be~\cite{Fischer:2018lme}
\be
\label{ew}
\imag g_R = -2.175 \times 10^{-3} \, .
  \en
We agree with the result in \cite{Arhrib:2016vts} but disagree with
\cite{GonzalezSprinberg:2011kx}.

A bound on $\imag g_R$ can be obtained by first calculating the contribution
of $\imag g_R$ to the structure function $D$. The result is
\be
\frac{D}{A^{(0)}}=-\frac{3\pi}{4}\frac{(1-x^2)}{(1+2x^2)}
\imag g_R
\en

A positivity bound on $\imag g_R$ can then be derived in analogy to the bound
on the
$T$--even structure
  function $C$ in Eq.~(\ref{boundC}) but now setting $\sin\phi=1$ and
  replacing $C$ by $D$. One then obtains the $O(\alpha_s)$
bound~\cite{Groote:2017pvc}
\be
-0.0420 \le \imag g_R \le 0.0420\,.
\en
For once, the $CP$--conserving electroweak absorptive contribution to
$\imag g_R=\,-2.175 \times 10^{-3}$ can be seen to easily satisfy the bound.

As concerns experiment the bound is tighter than the experimental bound
obtained by the ATLAS
collaboration~\cite{Aaboud:2017aqp}
\be
-0.18 \le \imag g_R \le 0.06 \, .
\en

We mention that the NLO electroweak corrections to the two polarized $T$--even
structure functions $B$ and $C$ have not been done.

  \section{The two-stage sequential polarized top quark decay $t(\uparrow) \to b + W^+
    (\to\ell^+ +\nu_\ell)$}
  \label{sec-seq}
\begin{figure}[h]
    \centering
\includegraphics[width=7cm,clip]{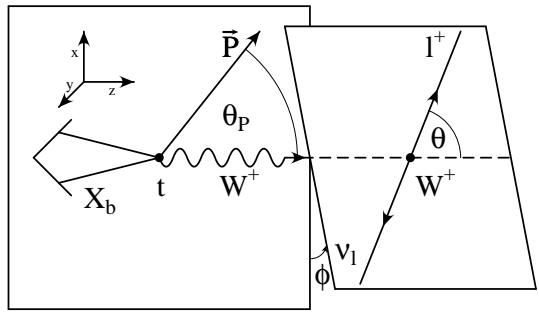}
\caption{Definition of the polar angles $\theta$ and $\theta_P$,
  and the azimuthal angle $\phi$ in the two-stage sequential two-body decay
  $t(\uparrow) \to X_b +W^+ (\to \ell^+ +\nu_\ell)$.}
\label{3angles}       
  \end{figure}

  The two-stage sequential two-body decay process $t(\uparrow) \to b + W^+
    (\to\ell^+ +\nu_\ell)$ is described by two polar angles
  $\theta$ and $\theta_P$, and the azimuthal angle $\phi$ as defined in
  Fig.~\ref{3angles}.

 The count of the independent structure functions is best done by
considering the independent double spin density matrix elements
$H_{\lambda_W\, \lambda'_W}^{\,\lambda^{\phantom x}_t\,\lambda'_t}$ of the
$W^+$  which form a hermitian $(3\times3)$ matrix 
\be
\bigg(H_{\lambda_W\, \lambda'_W}^{\,\lambda^{\phantom x}_t\,\lambda'_t}
  \bigg)^\dagger=
\bigg(H_{\lambda'_W\, \lambda_W}^{\,\lambda'_t\,\lambda_t}\bigg) \, .
\en 
  There are altogether ten independent double spin density matrix elements
\be
H_{++}^{++},\,H_{++}^{--},\,H_{--}^{++},\,H_{--}^{--},\,H_{00}^{++},\,%
H_{00}^{--},\,\real H_{+0}^{+-},\,\imag H_{+0}^{+-},\,\real H_{-0}^{-+},
\,\imag H_{-0}^{-+}
\en
out of which eight are T-even and two are T-odd structure functions. Compare
this to the three T-even and one T-odd structure functions
that describe the direct three-body decays of polarized top quarks discussed in
Sec.~\ref{sec-2}.

Let us
concentrate on the polar angle distribution which is obtained by
integrating over the azimuthal angle $\phi$. One obtains
\bea
    W(\theta,\,\theta_P) & =&   
    \frac{3}{8} \Big(1 \,+\, \cos \theta \Big)^2 \Big(T_+
    + T_+^P P_t \cos \theta_P\Big)  + 
    \frac{3}{8} \Big(1 - \cos \theta \Big)^2 \Big(T_-
    + T_-^P P_t \cos \theta_P\Big) \nn
    &+&\frac{3}{4} \sin^2 \theta \Big(L +L^P \, P_t \cos \theta_P \Big)
    \ena
    where $T_+,T_+^P,T_-,T_-^P,L,L^P$ are linear combinations of
    $H_{++}^{++},\,H_{++}^{--},\,H_{--}^{++},\,H_{--}^{--},\,H_{00}^{++},\,%
    H_{00}^{--}$.

    At LO and for $m_b=0$ one has
    $T_- = - T^P_-$, $L = + L^P$, $ T_+ = T_+^P=0 $, i.e. the $\cos \theta_P$
    dependence of the longitudinal and transverse-minus rates are given by
    $L:\, (1+P_t\cos\theta_P)$ and $T_-:\,(1-P_t\cos\theta_P)$. Similar
    to the discussion in Sec.~\ref{sec-2.2} one is precariously close to a
    violation
    of positivity when $P_t=1$. Using the $O(\alpha_s)$ results
    of~\cite{Fischer:1998gsa,Fischer:2001gp} we
    have checked that positivity is not spoiled at NLO QCD. In the same vein
    one can derive NLO bounds for the $T$--even and $T$--odd structure
    functions $\real H_{+0}^{+-},\,\imag H_{+0}^{+-},\,\real H_{-0}^{-+},
    \,\imag H_{-0}^{-+}$ which again are satisfied by the NLO QCD and the
    NLO electroweak results.

    We mention that we are in the process of calculating the NLO 
    electroweak corrections to the eight $T$--even structure
    functions of sequential polarized top quark decay~\cite{DFGK}.  
     
\subsection{NNLO QCD corrections to sequential polarized top quark decay}
As a last topic of this presentation we discuss the calculation of NNLO QCD
corrections to the eight $T$--even structure functions in sequential
polarized and unpolarized top quark decays for
which we have
obtained some partial results~\cite{Czarnecki:2010gb,Czarnecki:2018vwh}.
The main idea behind our approach has been laid down in the NNLO
calculation of the total rate in~\cite{Blokland:2004ye}. One converts a
two-scale problem $\Gamma(m_t,m_W)$ to a one scale 
problem $\Gamma(m_t)$ bei expanding in the ratio $x=m_W/m_t$ such that


 \be
 \Gamma(m_t,\,m_W) \to \Gamma(m_t, \, \sum a_i \,x^i )
 \en
 In practice we terminate the expansion at $i=10$. We found very
 satisfactory convergence of the expansion.

 The NNLO results are obtained from the absorptive parts of 36 $O(\alpha_s^2)$
 three-loop top quark self-energy diagrams which we denote by $\Sigma$.
 The unpolarized and polarized rates are then obtained from  the trace
  \be
  \Gamma+\Gamma^P=\frac{1}{m_t}\,\imag tr\,\Big\{(p\!\!\!/_t+m_t)
  (1+\gamma_5 s\!\!/_t^\ell)\,\Sigma \Big\}
  \en
  where $s_t^\ell$ is the longitudinal polarization four-vector of the top
  quark. One needs to avail of a covariant representation of $s_t^\ell$ which is
  given by
  \be
   s_t^{l, \mu}  =  \frac{1}{| \vec{q} \,|} 
   \Big( q^{\mu} - \frac{p_t \!\cdot\! q}{m_t^2} p_t^{\mu} \Big),
   \en
   The unwieldy denominator factor $| \vec q\,|$ comes in through the
   normalization condition $s_t \cdot s_t=-1$.
   Express $| \vec q \,|= \sqrt{q_0^2-q^2}= \sqrt{(pq/m_t)^2 -q^2}$ through the
   (unphysical) inverse propagator of the top quark $N=(p_t+q)^2-m_t^2$.
  Then expand in terms of inverse powers of $N$ up to the desired order
  \be
  \frac{1}{|\vec q\,|} = \frac{2 m_t}{N}
     \sum_{i=0}^\infty {2i \choose i} \left( \frac{2 q^2 N
    - q^4 + 4 m_t^2 q^2}{4\,N^{2}}
  \right)^i  \,.
\en
One can replace $q^2$ by $m_W^2$ everywhere since one is cutting through
the $W^+$--line when taking the absorptive parts of the three-loop diagrams.

   In this way we have calculated NNLO QCD results for the three helicity
   fractions ${\cal F}_{T_+}$, ${\cal F}_{T_-}$ and
   ${\cal F}_{L}$~\cite{Czarnecki:2010gb} and the polarized rate
   $\Gamma^P_{T_+ + T_- + L}$~\cite{Czarnecki:2018vwh}. One finds that
   the perturbation series' are well-behaved. There is no real
   obstacle but hard work and the handling of huge computer codes to calculate
   the NNLO corrections to the remaining structure
   functions.
  
   We have checked elements of our three-loop calculation by doing the
   corresponding NLO calculation involving the absorptive parts of four NLO
   two-loop top quark self energy diagrams. The results of the $x=m_W/m_t$
   expansion agrees with the corresponding expansion of the known
   closed-form NLO results~\cite{Fischer:1998gsa,Fischer:2001gp}.
   
   For example, when one expands up to $O(x^{10},x^{10}\ln x)$ the NLO expansion
   reads
\bea
\hat\Gamma_{(T_+ + T_- +L)}^{(1)}&=&C_F\Bigg[\frac54+\frac32x^2-6x^4
  +\frac{46}9x^6-\frac74x^8-\frac{49}{300}x^{10}
  +\strut\nonumber\\&&\strut\qquad
  -2(1-x^2)^2(1+2x^2)\zeta(2)
  +\left(3-\frac43x^2+\frac32x^4+\frac25x^6\right)x^4\ln x\Bigg]\,,\nonumber\\
\hat\Gamma_{(T_+ + T_- +L)^P}^{(1)}&=&C_F\Bigg[-\frac{15}4-\frac{17}8x^4
  -\frac{1324}{225}x^5-\frac{31}{36}x^6+\strut\nonumber\\&&\strut\qquad
  +\frac{48868}{11025}x^7-\frac{23}{288}x^8+\frac{884}{6615}x^9
  -\frac3{100}x^{10}+(1+4x^2)\zeta(2)\Bigg]\,.
\ena
Note that the expansion of the parity-conserving ($pc$) rate $(T_+ + T_- +L)$
involves only even
powers of $x$ while the expansion of the parity-violating $(pv)$ polarized rate
$(T_+ + T_- +L)^P$
involves even and odd powers of $x$. This pattern holds true for all
$pc$ and $pv$ $O(\alpha_s)$ and the known $O(\alpha^2_s)$ $T$-even structure functions for
which we lack a deep understanding. Our hope is that some chance reader of this
presentation can provide us with a solution to this empirical paradigma. 

\section{Acknowledgement}
\label{ackn}
J, G, K. is grateful to the organizers of the International Workshop on QCD,
theory and experiment (QCD$@$Work) for the invitation to 
the workshop and their hospitality in Matera. Special thanks go to Fulvia
de Fazio and Pietro Colangelo.
S.~G. is supported by
the Estonian Science Foundation under the grant No.~IUT2-27. S. G.\ acknowledges
the hospitality of the theory group THEP at the Institute of Physics at the
University of Mainz where part of this work was done and the support of the
Cluster of Excellence PRISMA at the
University of Mainz.

\end{document}